\documentclass[useAMS,usenatbib]{mn2e}

\usepackage{natbib}
\usepackage{graphicx}
\usepackage{multirow}
\usepackage{txfonts}

\title[Spatial and statistical properties of small-scale convective vortex-like motions]{Spatial distribution and statistical properties of small-scale convective vortex-like motions in a quiet Sun region}
\author[S. Vargas Dom\'inguez, J. Palacios, L. Balmaceda,  I. Cabello and  V. Domingo]{S. Vargas Dom\'inguez$^{1}$\thanks{E-mail:
svd@mssl.ucl.ac.uk}, J. Palacios$^{2}$, L. Balmaceda$^{3}$,  I. Cabello$^{2}$ and  V. Domingo$^{2}$\\
$^{1}$Mullard Space Science Laboratory, University College London, Holmbury St Mary, Dorking, Surrey, RH5 6NT, UK.\\
$^{2}$Image Processing Laboratory, University of Valencia, P.O. Box: 22085, E-46980 Paterna, Valencia, Spain.\\
$^{3}$Instituto de Ciencias Astron\'omicas, de la Tierra y del Espacio (ICATE), CONICET, Argentina.}
\begin{document}

\date{Accepted xxxx September xx. Received 2010 August xx; in original form 2010 August xx}

\pagerange{\pageref{firstpage}--\pageref{lastpage}} \pubyear{2011}

\maketitle

\label{firstpage}

\begin{abstract}
High-resolution observations of a quiet Sun internetwork region taken with the Solar 1-m Swedish Telescope in La Palma are analyzed.  
We determine the location of small-scale vortex motions in the solar photospheric region by computing the horizontal proper motions of small-scale structures on time series of images. These plasma convectively-driven swirl motions are associated to: (1) downdrafts (that have been commonly explained as corresponding to sites where the plasma is cooled down and hence returned to the interior below the visible photospheric level), and (2) horizontal velocity vectors converging into a central point. The sink cores are proved to be the final destination of passive floats tracing plasma flows towards the center of each vortex. We establish the occurrence of these events to be 1.4 $\times$ 10$^{-3}$ and 1.6 $\times$ 10$^{-3}$  vortices  Mm$^{-2}$ min$^{-1}$ respectively for two time series analyzed here.
\end{abstract}


\section{Introduction}
\label{S:1}
The solar photospheric plasma is in constant evolution and is driven by convective processes in many spatial and temporal scales. Granular convective motions are of great interest as this process of energy exchange affects the evolution of magnetic structures while embedded in the plasma and might end up changing the topology of small emerging magnetic loops. By using high-resolution time series at rapid cadences  we are nowadays able to register some of the finest and fastest scales of solar activity in an attempt to determine the Sun's structural building blocks.  Investigations aim to describe the configuration and interactions of these features (i.e. magnetic concentrations) on their way up from the photosphere to upper layers in order to account for a more complete picture of solar activity.

The quiet Sun has proved to be more dynamic and magnetically active than previously thought  \citep{dominguez2003}, involving the rapid evolution of bright points (hereafter BPs) mostly associated to strong magnetic fields (up to the order of a few kG). Depending on their strength, magnetic concentrations are affected by the convective motions, being dragged and reconfigured through, for instance, fragmentation and coalescence phenomena \citep{viticchie2009}. Moreover, small-scale and short-lived emerging magnetic loops represent a substantial part of the total magnetic flux \citep{martinez2009} and therefore are useful to understand how they interact with the photospheric plasma. It is also important to remark that granular convection has been proposed to be involved in the mechanism driving the production of an efficient turbulent dynamo responsible for the quiet Sun magnetism \citep{petrovay1993}.

Vortical motions in the quiet photosphere associated to intense downdrafts are commonly found in theoretical simulations \citep{spruit1990,zirker1993,stein1998}. Very recent simulations by \cite{shelyag2011} have also shown that large amount of vorticity in the photosphere is formed due to the interaction of plasma with the magnetic fields in the intergranular lane junctions. These types of vortex-like motions have been observationally reported at large spatial scales by \cite{brandt1988,title1992,attie2009}; at small scales by \cite{bonet2008}; and in the chromospheric quiet Sun by \cite{wedermeyer2009}. 

In the present work we concentrate on a quiet Sun internetwork region that has been target of a few recent studies describing the presence of convective driven vortex flows \citep{bonet2008,bonet2010}, the coverage of BPs \citep{sanchez2010}, and evidence of magnetic concentrations dragged by a swirl motion \citep{balmaceda2010}.  We use time series of G-band images recorded at high cadence in order to compute the proper motions of structures in the observed region. By generating the horizontal velocity maps we discriminate the regions displaying a pattern of converging flows into a central point of intense strong downdrafts where the plasma is returned to the solar interior. These events are then tracked within the field-of-view (hereafter FOV). We estimate their lifetime to be around 10 to 20 min. Our results evidence that the detected events can not be considered as homogeneously distributed over the FOV. We highlight the importance of understanding the effects of photospheric vortical motions on the configuration of magnetic fields in their way up to higher solar atmospheric  layers, that are yet to be found.

\begin{figure}
\includegraphics[angle=90,width=.65\linewidth]{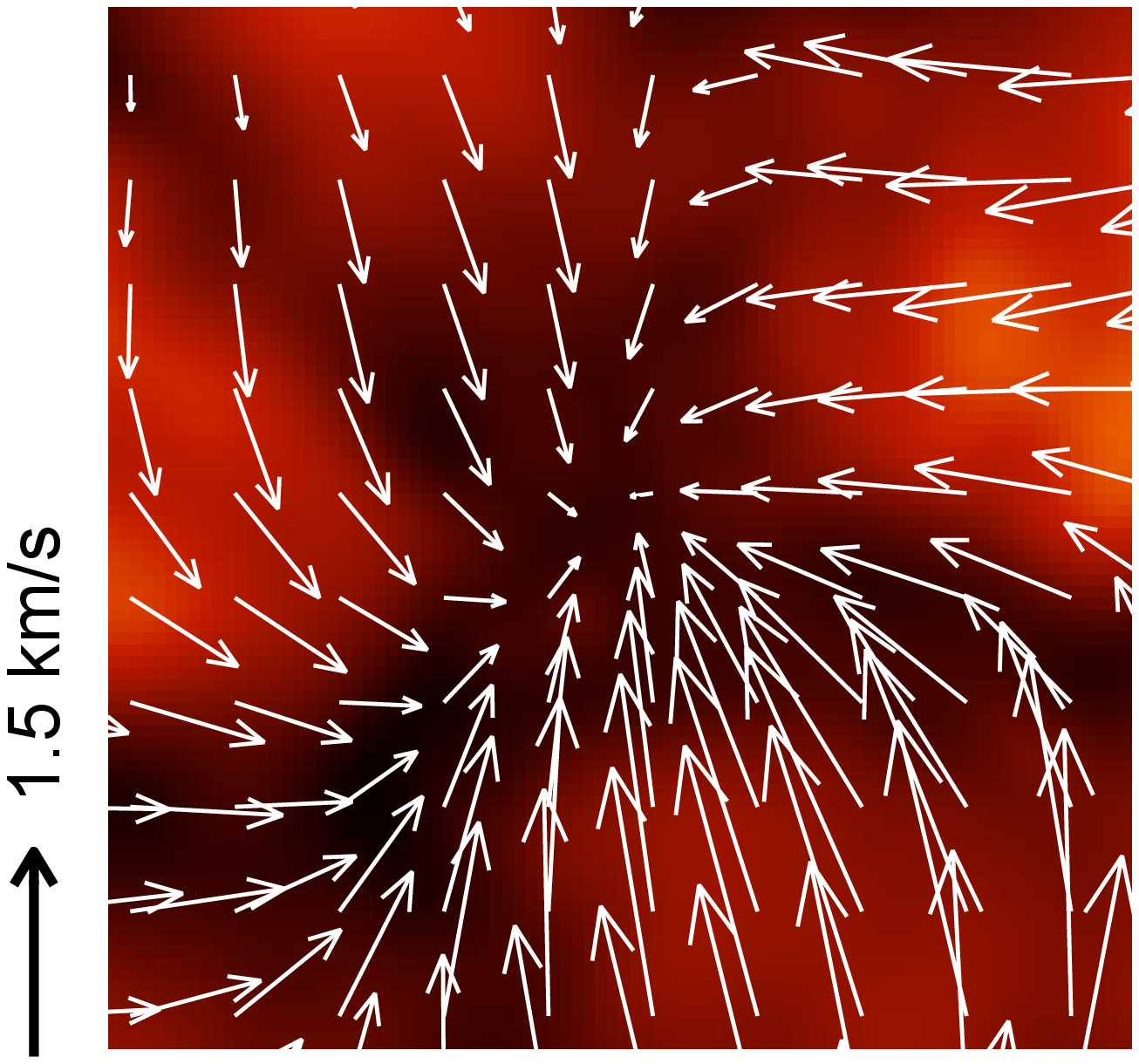}
\caption{Example of one of the detected events displaying horizontal flows converging into a central point. Arrows represent the horizontal velocity vectors inferred from LCT. The displayed swirl motion exhibits a counterclockwise sense of rotation. The background image represents the 20-min average image in false color with a FOV of 3$\arcsec \times$ 3$\arcsec$.}
\label{vortexsample}
\end{figure}

\begin{figure*}
\includegraphics[angle=90,width=.33\linewidth]{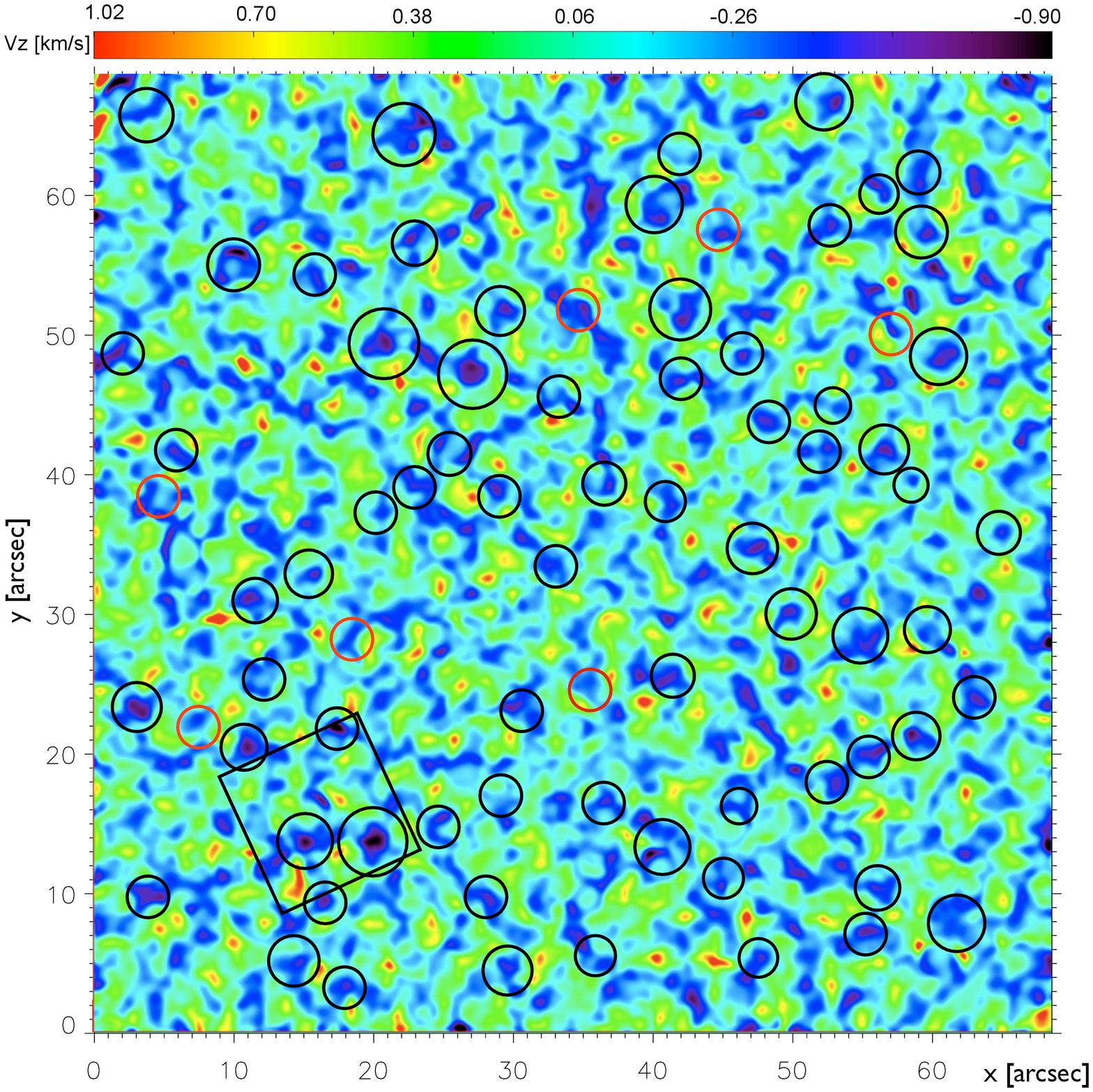}\\
 \includegraphics[angle=90,width=.33\linewidth]{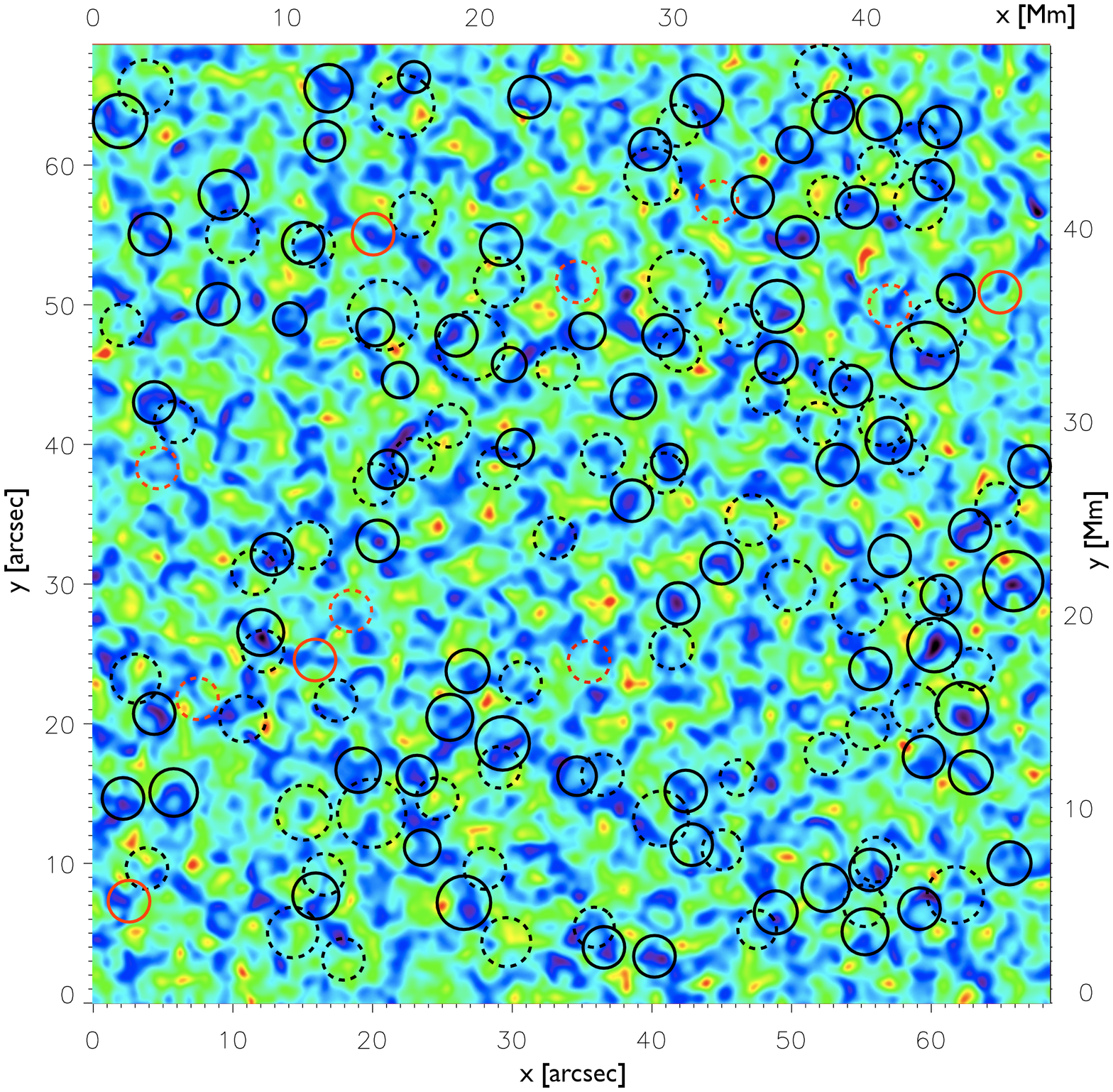}
\caption{Maps of vertical velocities derived from the horizontal velocity field using LCT analysis for the SST/G-band time series \emph{s1} (\emph{upper panel}) and \emph{s2}  (\emph{lower panel}). The same time coverage of 20 min is used in both time series. The black box represents the FOV analyzed by \citet{balmaceda2010} including a pair of strong swirl events.  Encircled in black are regions in which the horizontal velocities are converging towards a central point as inferred from the flow maps. The size of circles indicates the area coverage of the corresponding swirl event. Red circles denote peculiar regions with vertical velocities that do not conform vortical flows. Dashed circumferences in the lower panel show the location of the events in the upper panel for comparison.}
\label{verticalvel}
\end{figure*}

\begin{figure*}
\includegraphics[angle=0,width=.85\linewidth]{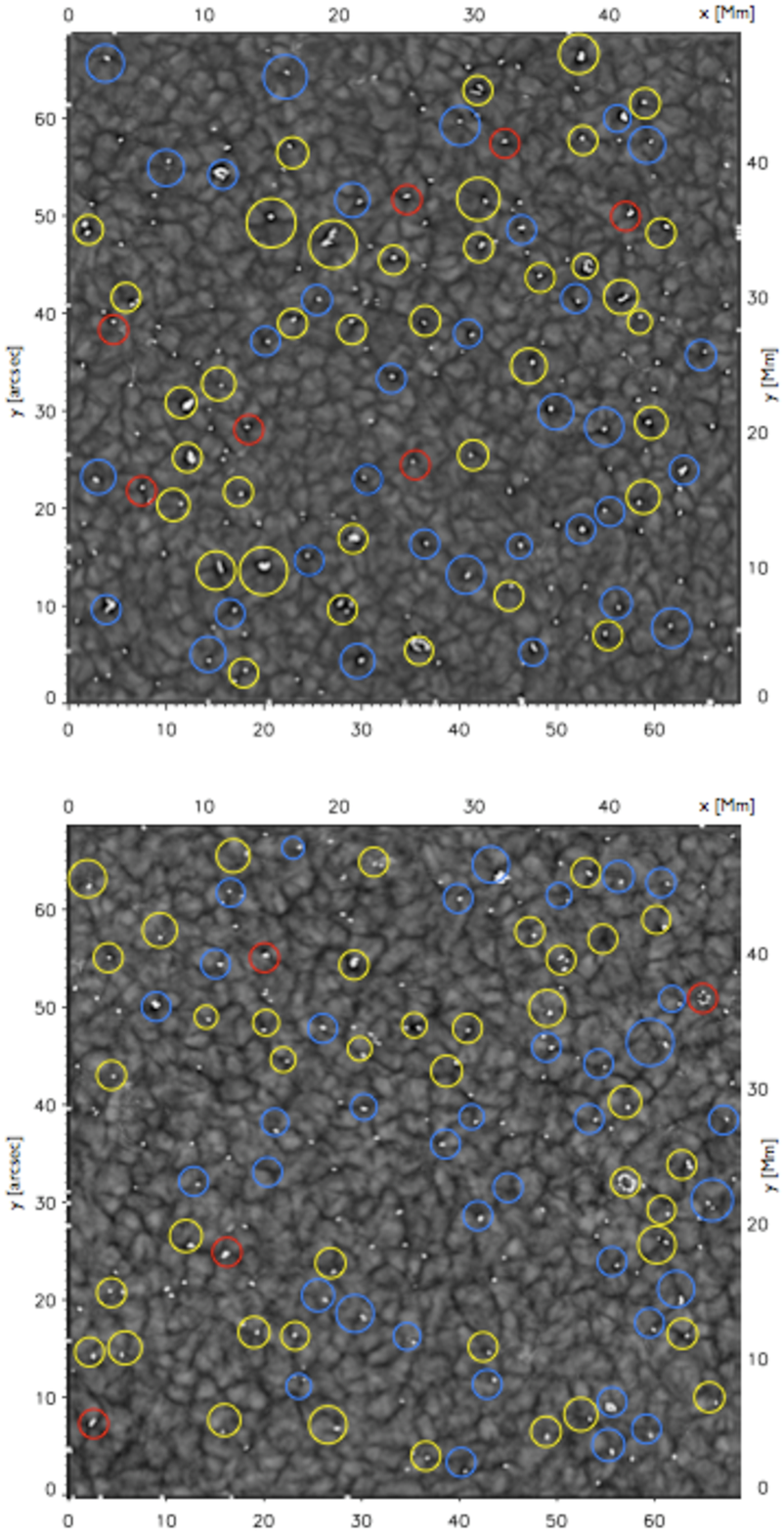}
\caption{The images show the final destination of passive tracers (\emph{in white}) driven by the computed mean field of horizontal velocities. The background represents the 
high-contrast image smoothed by the 20-min average in every case. Locations of the swirl events detected in Fig.~\ref{verticalvel} (in \emph{black circles}) are encircled in \emph{blue/yellow} denoting  clockwise/counterclockwise sense of rotation, respectively. Red circles are regions in which there is a strong concentration of \emph{floats}, also plotted in Fig.~\ref{verticalvel}.}
\label{corchos}
\end{figure*}

\section{Observations and data preparation}
\label{S:2}

The data were acquired during a particular observing run on 29 September 2007 with the Swedish 1-m Solar Telescope \citep[SST,][]{scharmer2003} in La Palma, Canary Islands, as part of a long international campaign with joint observations making use of other solar facilities at the Canary Islands Observatories. The region of interest corresponds to a quiet Sun area close to the solar disc centre ($\mu$=0.99).  Images in G-band ($\lambda 430.56$ nm) were recorded at a fast cadence and the \emph{Multi-Frame Blind-Deconvolution} \citep[MFBD,][]{lofdahl2002} restoration technique was applied to correct the images from the aberrations induced by the turbulent atmospheric medium that affect their quality. The restored images with an effective FOV of  $69 \times 69$ square arcsec and spatial sampling of $0\farcs034$/pix were corrected with the standard procedures of flat-fielding, dark current substraction, elimination of spurious pixels and borders, and were grouped in two continuous time series of images, \emph{s1}: 08:47 to 09:07 UT, and \emph{s2}: 09:14 to 09:46 UT, with 15s cadence. The $\sim$7-min gap between \emph{s1} and \emph{s2} corresponds to poor quality images that had to be discarded due to bad observing conditions. The final steps included the compensation for diurnal field rotation, destretching and subsonic filtering to eliminate residual jittering. More details on the data preparation can be found in \cite{balmaceda2010}.

\section{Sinks and converging flows} 
\label{S:3}

Our analysis is based on the widely used local correlation tracking techniques \citep[LCT,][]{november1988}, implemented by \cite{molowny1994}. We compute the horizontal proper motions of structures using a Gaussian tracking window with a Full-Width at Half-Maximum (FWHM) of 1$\farcs$0, to generate maps of horizontal velocity (flow maps). The same procedure was used by \cite{balmaceda2010} on the same data for the analysis of a fraction of the whole FOV  (10$\arcsec$ $\times$ 10$\arcsec$) displaying a region of strong negative divergence. These authors found converging horizontal flows, i.e  the velocity arrows pointing to a common destination being the location of the sink. 

In this work we present the results obtained from the LCT analysis for series \emph{s1} and \emph{s2} over the complete FOV for 20-min intervals. From the horizontal velocity maps we compute the divergence 
field defined as: $\nabla \vec v=\frac{\partial \vec v_{x}}{\partial x}+\frac{\partial \vec v_{y}}{\partial y}$, where $ v_{x}$ and  $ v_{y}$ are the corresponding $x$ and $y$ components of the horizontal velocity vector.

Vertical velocities are inferred after multiplying the horizontal flow divergence by the so-called scale height  of the flux mass ($h_m=150$~km), following \cite{november1989}. Note that the term divergence refers to the divergence of the horizontal velocities. For a detailed derivation, physical explanation and validity of this relation, we refer the reader to \cite{marquez2006}. Resulting vertical velocities are manifestly conditioned by the LCT average and therefore can not be directly compared to  Doppler velocities. An upcoming work attempts to establish margins for the comparison of the two. In our data the velocity values as corresponding to a 20-min average range between 1.8 (upflows) and -1.2~km~s$^{-1}$ (downflows). The overall distribution of horizontal velocities of one of these events is shown in Fig.~\ref{vortexsample}. There is a clear trend in the direction of the velocity vectors, pointing towards the centre of the image (\emph{draining point}) and forming a swirl motion with a counterclockwise sense of rotation. The background image is the G-band intensity average over the 20-min interval. The centre coincides with a dark structure formed by the junction of various intergranular lanes. 

Figure~\ref{verticalvel} displays the vertical velocity maps for \emph{s1} (\emph{upper panel}) and  \emph{s2} (\emph{lower panel}) respectively, using a common time coverage of 20 min for both time series (i.e. the total duration of \emph{s1} and the first 20 min of \emph{s2}) in order for the results to be comparable. Regions showing intense upflows denoted in different shades of red color correspond to intensive and recurrent activity from exploding granules with velocities of $\sim$1~km~s$^{-1}$, whereas strong sinks  with negative, i.e. entering the plane of the figure, vertical velocities of magnitude $\sim$0.9~km~s$^{-1}$ are observed as dark features. The intergranular lanes are generally traced by the elongated structures in blue with downflows (-0.2~km~s$^{-1}$). The black box in the upper panel represents the FOV studied by \cite{balmaceda2010}. Encircled in black are the locations of the detected vortical motions (also referred to as events) from the horizontal flow pattern. The selection was performed by visual inspection, considering a vortex where velocity vectors in the flow map was converging and changing direction defining a swirl. The area coverage of the swirl events are assumed to be circular and drawn as circles in the maps following the mentioned visual criteria. In all detected cases the events are located in regions with strong downflows  ($v_z$ $\sim$ -0.8~km~s$^{-1}$).  Note that $v_z$ represents the averaged $v_{LOS}$ over the time elapsed. The presence of all the events in at least two of the consecutive flow maps over 5-min intervals suggest that they remain coherent in a range between 10 and 20 min.  The positions of the events change from the first to the second time series after the 7-min gap in the observing run (\emph{dotted circles} in the \emph{lower panel} in Fig.~\ref{verticalvel} show the location of the events in the upper panel in the same figure for comparison.) These events do not seem to be homogeneously distributed over the FOV but rather grouped in certain locations resembling the mesogranular and supergranular patterns.

Figure~\ref{corchos} shows the final destination of passive tracers or \emph{floats} (\emph{white dots} in the figure) driven by the computed mean field of horizontal velocities, as commonly used to trace the evolution of plasma motions \citep[see for instance][]{yi1992,marquez2006}. The background represents the averaged image in every case. The position of the events detected in Fig.~\ref{verticalvel} are this time encircled in \emph{blue}/\emph{yellow} denoting clockwise/counterclockwise sense of rotation. For both time series we found a larger number of counterclockwise motions, though the difference is not very significant.  In general these events are evidenced by a larger concentration of \emph{floats} with the exception of a very few cases in the second time series (\emph{lower panel}) in which no sign of \emph{floats} is detected though the areas are characterized by a junction of somewhat dark intergranular lanes.  The distribution of vortices shows therefore a prominent  correspondence with the location of  intergranular lanes. The \emph{red} circles in Fig.~\ref{corchos} are regions in which there is a strong concentration of \emph{float} tracers, forming in some cases ring-shaped features. Although some of these regions display converging flows and correspond to downflows, they are not linked to very strong downflows when plotted over the image of vertical velocities (see \emph{red circles} in Fig.~\ref{verticalvel}). Their location also change from the first to the second time series (e.g. compare the distribution of \emph{red} and \emph{dotted-red} circles in the lower panel of Fig.~\ref{verticalvel}).  They do not represent though the most significant part of the detected events but $\sim$10\% in both series, respectively.  We have calculated the occurrence of the detected vortex-type events in the FOV obtaining 1.46~$\times$~10$^{-2}$ and  1.09~$\times$~10$^{-2}$ ~vortices ~Mm$^{-2}$ for the SST/G-band series \emph{s1} and  \emph{s2}, respectively (the values increase up to  1.66~$\times$~10$^{-2}$ and  1.25~$\times$~10$^{-2}$~vortices ~Mm$^{-2}$ if including the events not associated to very strong downflows and encircled in red in Fig.~\ref{corchos}.)

\section{Statistics of swirl motions}
\label{S:5}

As discussed in the previous section we have computed horizontal velocities and their corresponding divergences from G-band observations. With the location of the detected vortices (circles in Fig.~\ref{corchos}) we define a binary mask used to compute some statistical properties in vortex-type areas (i.e., a mask of vortices).  The vorticity is a vectorial magnitude whose direction is perpendicular to the $XY$ plane.  Using the equation: $\nabla \times \vec v=\frac{\partial \vec v_{y}}{\partial x}-\frac{\partial \vec v_{x}}{\partial y}$, we compute vorticities of the horizontal flow, with $v_z$  not included in the computation. Previous works have already studied vorticity in the solar photosphere from observational \citep{wang1995,bonet2010} and theoretical \citep{stein1998,shelyag2011} approaches. 

\begin{figure*}
\includegraphics[angle=90,width=.88\linewidth]{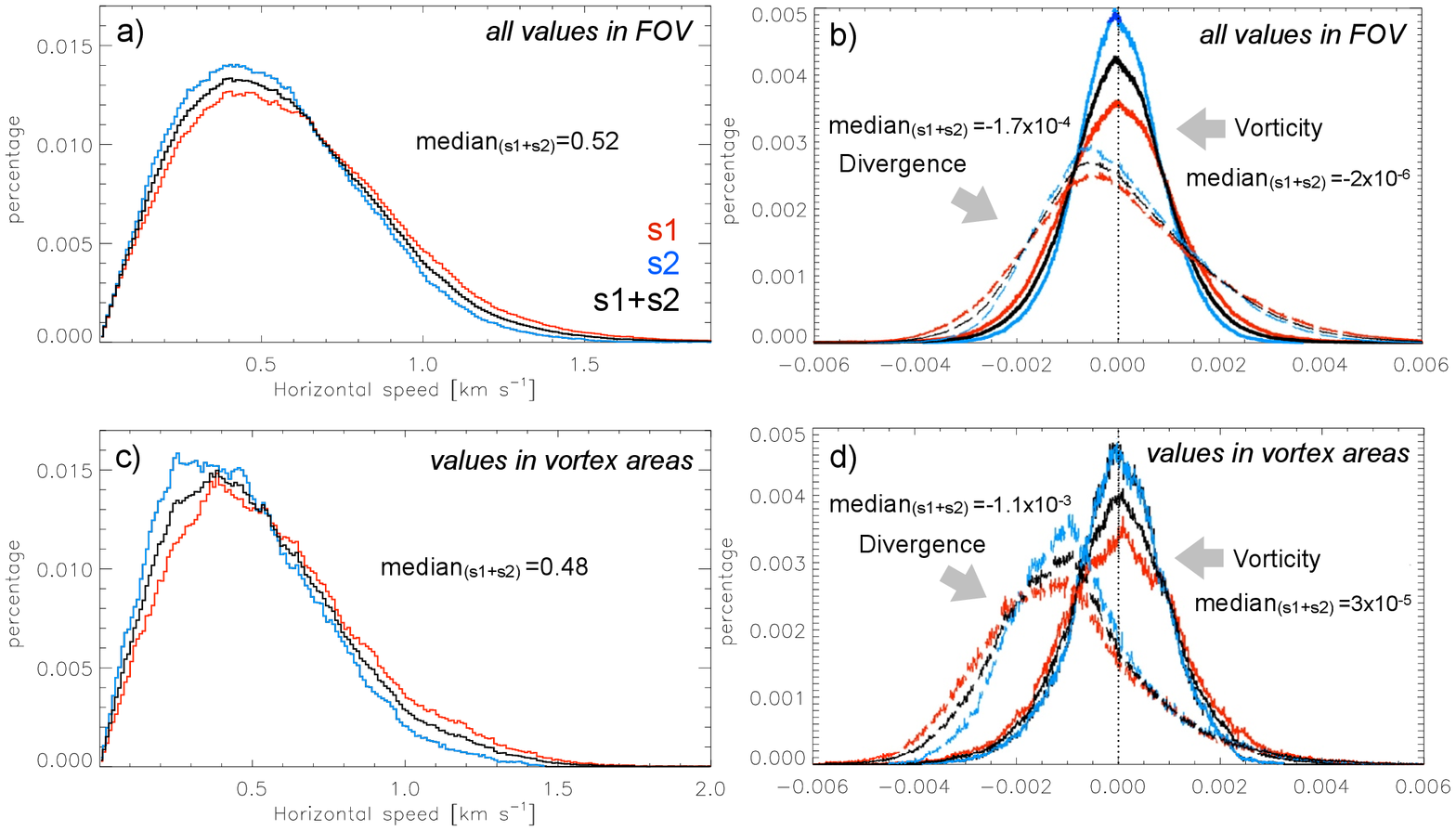}
\caption{\emph{Panels} a) \emph{and} c): Distribution of horizontal speeds. \emph{Panels} b) \emph{and} d): Histograms of the values of divergence (\emph{dashed lines}) and vorticity (\emph{solid lines}). Upper panels plot values in all the FOV whereas lower ones account for values within areas displaying vortices.  All plots are computed using values for series s1 (\emph{in red}),  s2 (\emph{in blue}) and combined values for both time series  (\emph{in black}). Corresponding most probable values for the combined case are shown in the panels.}
\label{histogramas}
\end{figure*}

Figure~\ref{histogramas} (upper panels) shows histograms for the whole FOV of  $\sim50 \times 50$ square Mm. Panel a) displays independent results for series s1 (\emph{red}) and s2 (\emph{blue}) and total values for both series  (\emph{black}), resulting in a Maxwellian distribution of horizontal velocities, with a most probable value of 0.52~km~s$^{-1}$. Histograms in panel c) display a similar distribution of speeds using exclusively values within the mask of vortices. The most probable speed reduces to 0.48~km~s$^{-1}$. Median values combining s1 and s2, referred to as median$_{(s1+s2)}$, are shown in each panel in Figure~\ref{histogramas}. Independent median values of horizontal speed for s1/s2 are 0.50/0.55~km~s$^{-1}$  in Figure~\ref{histogramas}a and 0.51/0.44 ~km~s$^{-1}$  in Figure~\ref{histogramas}c. Large horizontal velocities in the FOV are predominantly coming from recurrent exploding events (at the mesogranular scale) with large positive divergence values and hence the mean speed within vortex areas is expected to moderate as corresponding to negative divergences (convergences).\\

\begin{figure*}
\includegraphics[angle=-90,width=.88\linewidth]{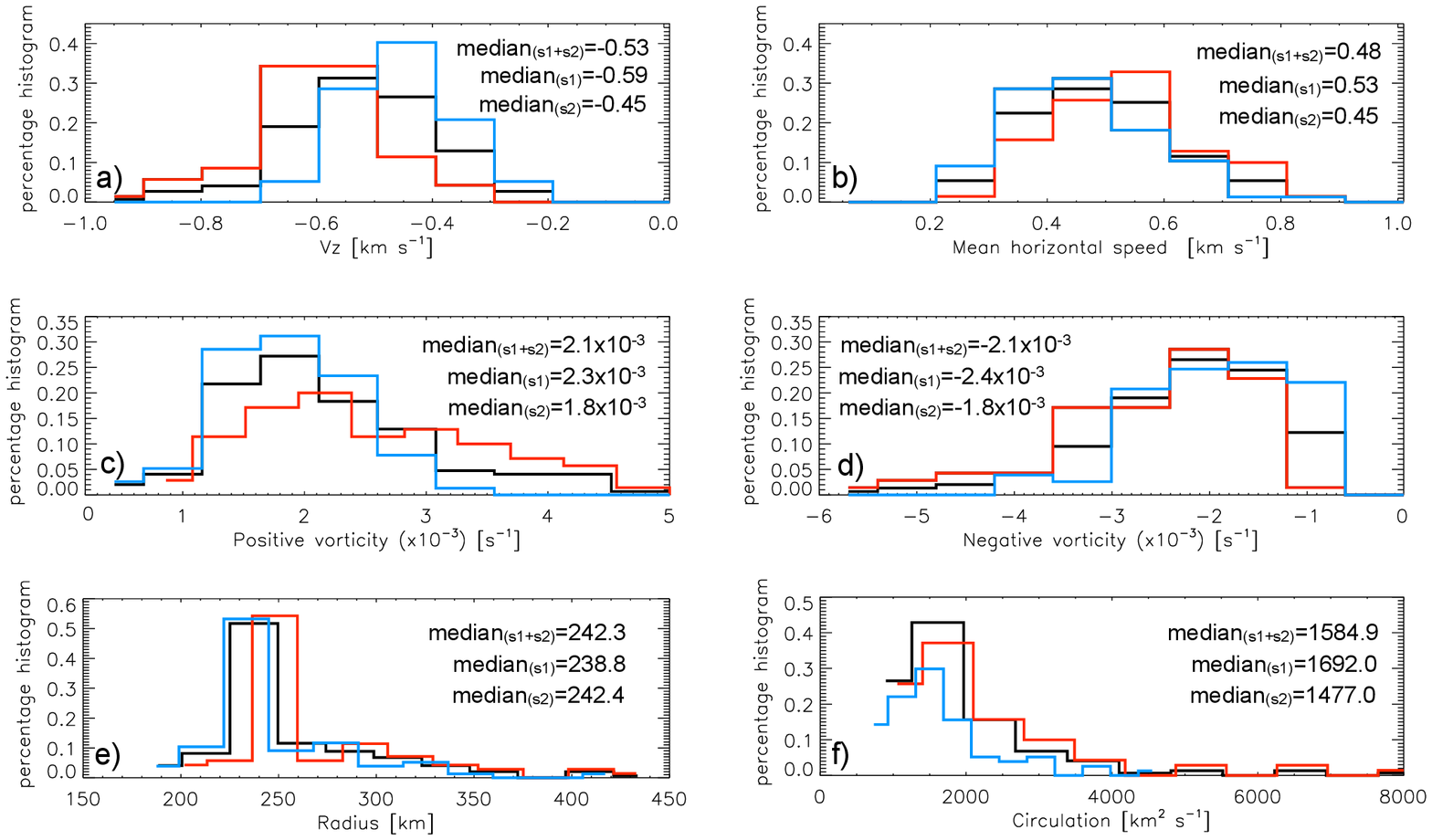}
\caption{Histograms for data within the mask of vortices showing mean values of:  a) vertical velocity, b)  horizontal speed, c) positive vorticity, d) negative vorticity, e) radius of vortices and f) circulation, computed
for the vortices in series s1 (\emph{in red}),  s2 (\emph{in blue}) and combined values for both time series  (\emph{in black}). Corresponding most probable values for s1, s2 and for the combined case (s1+s2) are shown in the panels.}
\label{vortexhist}
\end{figure*}

\begin{figure*}
\includegraphics[angle=-90,width=.9\linewidth]{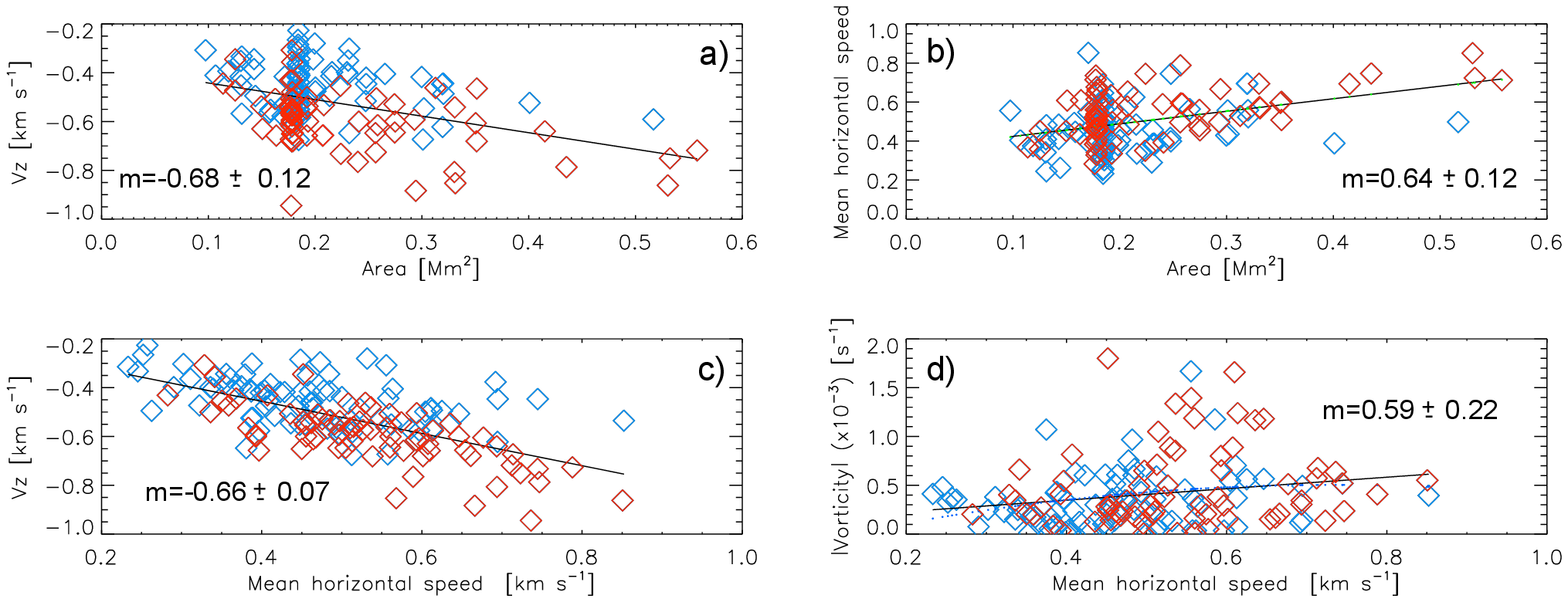}
\caption{Scatter plots of properties characterizing the detected vortices.  Diamonds represent the values for s1 (\emph{red}) and s2 (\emph{blue}) . Black lines are the result of linear fits applied over the data with their corresponding regression coefficients and errors shown.}
\label{vortexscatter}
\end{figure*}

Same as with horizontal velocities and divergences, we have also plotted vorticity values over the whole FOV and in the mask of vortices (i.e. in the areas inside circles in Figs.~\ref{verticalvel} and \ref{corchos}). Figures~\ref{histogramas}b) and \ref{histogramas}d) show the distribution of divergence (\emph{dashed lines}) and vorticity (\emph{solid lines}).  Independent results are again plotted for series s1 (\emph{red}) and s2 (\emph{blue}) and total values for both series  (\emph{black}). When considering the total FOV, the divergence distribution is centered at a slightly negative value indicating a preference for converging flows probably located at the edges of supergranular cells. This behaviour is more evident when considering vortex areas only (\emph{panel d}). The vorticity distribution for the whole FOV shows a slight trend for clockwise sense of rotation. Areas presenting vortices instead show a counterclockwise trend. 

Figure~\ref{vortexhist} plots a series of histograms computed using exclusively average values within the mask of vortices (i.e. mean values for each vortex). Vortices detected in series s1 and s2 respectively show different vertical velocities $vz$, with more intense downflows observed in the second one. From the histograms in Fig.~\ref{vortexhist}a, most of the vortices are characterized by velocities of -0.4~km~s$^{-1}$ and -0.6~km~s$^{-1}$ for s1 and s2, respectively. It is important to bear in mind that $vz$ is always negative since all vortices are observed in downflow areas, i.e. areas where the divergence is negative, with $vz$ being proportional to divergence times the scale height. Fig.~\ref{vortexhist}b shows the distribution of the same quantities from Fig.~\ref{histogramas}a and \ref{histogramas}c but considering the mean values of speed in areas displaying vortices. The results are similar. In Fig.~\ref{vortexhist}c and Fig.~\ref{vortexhist}d the distributions for positive and negative vorticities are shown. The medians for the counterclockwise and clockwise sense are comparable, indicating an apparent non-preferred sense of motion though in Section~\ref{S:3} we found counterclockwise events as being slightly more numerous. Fig.~\ref{vortexhist}e reveals a strong peak in the values between 240 and 250 km for the mean radius. It is worth noting the small size of the vortices that remain stable during the two series since we would expect larger sizes i.e. of mesogranular scales, considering the 20-minute average used for the LCT computation. In Fig.~\ref{vortexhist}f we include the circulation, defined for each vortex as the mean absolute vorticity times their corresponding area. A peak value of 1584.9~km$^2$~s$^{-1}$ is clearly enhanced in the plots.

Figure~\ref{vortexscatter} displays scatter plots comparing different properties that characterize the vortices for series s1 (\emph{blue}) and s2  (\emph{red}). Linear fits are shown in \emph{black} for the combined values of both series. For all panels regressions were also separately computed for each series (not shown in the plots) with the resulting fits matching well between the two series. Panels a) and b) show the dependence of the vortex area and its corresponding velocities: larger vortices present higher horizontal and vertical speeds. This behavior is also clear in panel c). Larger scattering is observed in panel d), where the dependence of mean horizontal speeds with vorticities is shown. However, the linear fit shows a trend of slightly higher vorticities for larger speeds.

\section{Discussion}
\label{S:4}

Recent advances in solar instrumentations from ground-based and satellite telescopes are providing us with unprecedented high-resolution data allowing the study of the dynamics of the finest solar structures. The evolution of the solar granulation pattern at the photospheric level is of great interest due to the interaction between plasma and magnetic concentrations. In this scenario vortices, formed in junctions of multiple intergranular lanes, are places in which the concentration of these magnetic features are expected to be favoured and therefore play an important role in the dynamics of the quiet Sun \citep{kitiashvili2010}.  

We search for the location of small-scale swirls by tracking plasma motions in the same region where \cite{bonet2008} detected small whirpools by following the trajectories of BPs  being swallowed by them.  From 20-min average flow maps we identify events displaying a converging pattern of horizontal velocity vectors towards a central point that correspond to the strong sinks as initially predicted by the numerical simulations. These events are in all cases detected in the vertices of multiple granules along very intense intergranular dark lanes. This is clearly seen on the average image for the duration of the time series where the low intensity junctions are enhanced (in Fig.~\ref{corchos}).  Data from two time series and common examination allow the comparison of results. A total of 70 in s1 and 77 in s2 vortices have been detected, resulting in a density of 2.8 $\times$ 10$^{-2}$~vortices ~Mm$^{-2}$ and  3.1 $\times$ 10$^{-2}$~vortices ~Mm$^{-2}$ for s1 and s2 respectively.

Averaging over the 20-min window we obtained space-time-density values of 1.4 $\times$ 10$^{-3}$~vortices ~Mm$^{-2}$~min$^{-1}$ and 1.6 $\times$ 10$^{-3}$~vortices ~Mm$^{-2}$~min$^{-1}$ for s1 and s2 respectively. Values are comparable to the number obtained by  \citet{bonet2008} of 1.8 $\times$ 10$^{-3}$ vortices  Mm$^{-2}$ min$^{-1}$ and lower than the space-time-density of 3.1 $\times$ 10$^{-3}$~vortices ~Mm$^{-2}$~min$^{-1}$ found by \cite{bonet2010}. It is necessary to bear in mind that the results from our flow fields are smoothed by the 20-min averages and the size of the employed tracking window. Many short-living vortical motions are likely diluted by the LCT temporal average.

BP«s are distributed all over the FOV in this region as found by \cite{sanchez2010} from which some describe spiral trajectories whilst being engulfed by downdrafts that \cite{bonet2008} described as convectively driven vortex flows. We have compared the location of our detected vortices with the events discovered  by \cite{bonet2008} in the same solar region. As a result, we have found 68\% of coincidences but the remaining 32\% corresponds to cases in the near vicinity of our vortices  (i.e. less than 2$\arcsec$ from the edge of the circular area enclosing the events in Fig.~\ref{verticalvel} and Fig.~\ref{corchos}.

We find for the series s1 and s2 counterclockwise/clockwise sense of rotation frequency value ratios of 53/47 and 52/48 \%, respectively, which in view of the number of studied events shows no significant difference with equally probable sense of rotation, in agreement with \cite{bonet2008}, though \cite{bonet2010} find a significant preference in favor of counterclockwise sense of rotation. The latter explain the different results assuming that the vortex rotation is influenced by the latitudinal solar differential rotation, as the observations in \cite{bonet2010} were done at mid solar latitude, while the observations in \cite{bonet2008} and, of course, in this paper were done at the equator, where the influence of differential rotation is negligible.

In terms of dimensions our detected vortices coincide with the values in \cite{bonet2010} of less than 500 km of radius. The majority of our detected vortices exhibit a radius of  241 $\pm$ 25 km. Some vortices, however, have a radius over 400 km that might not correspond to the main vortical scale detected but to signatures of larger scales of vortical motions.
For the circulation we also found a pronounced  peak value, at 1585~km$^2$~s$^{-1}$. This value is nearly half  the 4000~km$^2$~s$^{-1}$ for the vortex presented by \cite{brandt1988} which covers an area of about 5 granules  and hence corresponds to a much larger mesogranular scale.

Strong interacting vortices may play a role in excitation of solar acoustic oscillations, as shown by  \cite{kitiashvili2010}.  In particular these authors propose that high-speed vortices
can attract and capture other vortices of opposite sense of rotation and their interaction might lead to their partial annihilation. In our sample of vortices we find several examples with these characteristics, though a deep analysis of the interaction of opposite-sign vorticities is out of the scope of the present work.

More observational evidence from spectropolarimeric data would be required to complete the analysis and determine up to what extent the magnetic concentrations in the nearest vicinity of the detected swirl events are affected by the appearance of the vortex. An example of magnetic features appearing as being dragged by this type of convective motions was presented by \cite{balmaceda2010} in one of the vortices within the FOV. This type of vortical convective flows are thought to contribute gathering magnetic fragments thus amplifying weak magnetic fields near the surface and representing an important mechanism for the formation of more stable features \citep{danilovic2010}. Exhaustive analysis of the intensification of magnetic field in vortex-like regions and its morphology at different solar atmospheric layers will also contribute with precise inputs to improve numerical simulations development.

\section*{Acknowledgments}
SVD acknowledges support from STFC. JP, VD and IC acknowledge funding from the Spanish grant BES-2007-16584 on the projects ESP2006-13030-C06-04 and AYA2009-14105-C06, including European FEDER funds.  IC acknowledges funding from project ESP-2003-07735-C04-04 on grant BES-2004-4372. The Swedish 1-m Solar Telescope is operated on the island of La Palma by the Institute of Solar Physics of the Royal Swedish Academy of Sciences in the Spanish Observatorio del Roque de los Muchachos of the Instituto de Astrof\'isica de Canarias, and programa de acceso a Grandes Instalaciones of the Spanish Science Ministry and IAC. We thank the scientist of the Hinode team for the operation of the instruments. Hinode is a Japanese mission developed and launched by ISAS/JAXA, with NAOJ as domestic partner and NASA and STFC (UK) as interna- tional partners. It is operated by these agencies in co-operation with ESA and NSC (Norway).

\label{lastpage}

\end{document}